\documentclass[runningheads,a4paper]{llncs}

\usepackage{amssymb}
\usepackage{graphicx}
\usepackage{algorithm}
\usepackage{algorithmic}
\usepackage{amsmath}
\usepackage{paralist}

\usepackage{url}
\newcommand{\keywords}[1]{\par\addvspace\baselineskip
\noindent\keywordname\enspace\ignorespaces#1}

\begin{document}

\mainmatter  

\title{Maximizing Revenues for Online-Dial-a-Ride}


%
%
%

%
%

\author{Ananya Christman\inst{1} \and William Forcier\inst{2}}

\institute{Middlebury College, Middlebury VT 05753\\
\email{achristman@middlebury.edu},
\and
Lake Forest College, Lake Forest IL 60045\\
\email{forciwc@lakeforest.edu}}

\maketitle

\begin{abstract}
\emph{
In the classic Dial-a-Ride Problem, a server travels in some metric space to serve requests for rides. Each request has a source, destination, and release time. We study a variation of this problem where each request also has a revenue that is earned if the request is satisfied. The goal is to serve requests within a time limit such that the total revenue is maximized. We first prove that the version of this problem where edges in the input graph have varying weights is NP-complete. We also prove that no algorithm can be competitive for this problem. We therefore consider the version where edges in the graph have unit weight and develop a 2-competitive algorithm for this problem.
}
\keywords{online algorithms, dial-a-ride, competitive analysis, graphs}
\end{abstract}

\section{Introduction}

In the Dial-a-Ride Problem (DARP), a server travels in some metric space to serve requests for rides. The server starts at a designated location of the space, the \textit{origin}. Each request specifies a \textit{source}, which is the the start location of the ride, a \textit{destination}, which is the end location, and the release time of the request, which is the earliest time the request may be served. The objective is to route the server through the metric space so as to meet some optimality criterion. In the On-Line-Dial-a-Ride Problem (OLDARP), the requests are issued dynamically to the server and the server is unaware of future requests. Therefore, for each request, the server must decide whether to serve the request and at what time. In many cases preemption is not allowed, so every request that the server decides to serve must be served until completion.

In the version of OLDARP that we consider, each request also has an associated revenue, which is the amount earned by the server for serving the request, and there is a global time limit $T$ such that requests must be served before time $T$. The goal is to serve requests within the time limit so as to maximize the total revenue. We assume preemption is not allowed, therefore serving a particular request may prevent the server from serving another request with a higher revenue.

On-Line Dial-a-Ride Problems have many practical applications such as vehicle routing, internet Quality of Service (QoS), and combatting terrorism. Vehicle routing applications include door-to-door transport services for elderly, disabled, or ill patients. Nodes represent patients and revenues represent the priority level for each request.  For QoS applications, internet service requests are issued to service providers who must decide which requests to serve to maximize total revenue. For applications related to combatting terrorism, the server may be robot that detects IEDs (Improvised Explosive Devices). The source-destination pair of a request may represent a city street that should be investigated and the revenue may represent the priority level of the request.

\section{Related Work}

Several variations of the On-Line Dial-a-Ride Problem have been studied in the past. In the general version of the problem each request has source, destination, unit load, and release date. A server with capacity $Q$ starts at an origin location and serves a request by picking up at the source and delivering at the destination. Requests arrive online to a server and can be served no sooner than the release date. Various objective functions have been investigated for this problem.

The authors of~\cite{stougie} also consider the unit metric space with two different objectives. One is to minimize the time the last destination is served (also known as \textit{makespan}); the other is to minimize the sum of \textit{completion times}, i.e. the time a request is served (also known as \textit{latency}).  They show that any deterministic algorithm must have competitive ratio of at least 2 independent of the server capacity $Q$. They also give a 2-competitive algorithm for $Q= \infty$.
For the objective of minimizing latency, they prove that any algorithm must have a competitive ration of at least $1 + \sqrt{2}$.

The work in~\cite{lipmann} considers a modified version of OLDARP where at the release time of a request only the pickup location is revealed. The delivery location is revealed only when the pickup operation is performed. Such a setting is appropriate for applications such as elevator scheduling or ride scheduling for taxis. The authors proved that when preemption is allowed (i.e. the server is allowed to halt a ride at any time and possibly proceed with it later) any deterministic algorithm must have competitive ratio of at least 3. The also give a 3-competitive algorithm to solve this problem.

The authors of~\cite{ausiello} consider a version of the problem where each request consists of a single location and a release date. This is referred to as the On-Line Traveling Salesman Problem (OLTSP). The server starts at an origin and must decide which requests to serve to minimize the latency. The authors study this problem on the Euclidean space and prove that no online algorithm can be better than 2-competitive. They give a 2.5-competitive non-polynomial time online algorithm and a 3-competitive polynomial time algorithm to solve this problem.

The authors of~\cite{ausiello2} also aim to minimize completion times for OLTSP but consider an asymmetric network where the distance from one point to another may differ in the inverse direction. They consider two versions of the problem: \textit{homing} where the server is required to finish at the origin, and \textit{nomadic} where there is no such requirement. They provide a non-polynomial $\frac{3+\sqrt{5}}{2}$-competitive algorithm for the nomadic version and prove that no competitive online algorithm can exist for the nomadic version.

The work in~\cite{jaillet2} considers a variation of OLTSP where each request also has a penalty (incurred if the request is rejected). The goal is to minimize the time to serve all accepted requests plus the sum of the penalties associated with the rejected requests. For the setting where the server can decide to accept/reject a request any time after the request's release date, the authors give a 2-competitive algorithm to solve the problem on the real line and a 2.28-competitive algorithm on a general metric space.

The authors of~\cite{krumke} studied both OLDARP and OLTSP for the uniform metric space. Their objective is to minimize the maximum \textit{flow time}, the difference between a request's release and service times. They prove that no competitive algorithm exists for OLDARP and give a 2-competitive algorithm to solve OLTSP.

Our version of OLDARP differs from previous research in that (1) each request is associated with a revenue (earned if the request is served) and (2) there is a time limit within which the server must complete all accepted requests.


\section{Problem Statement}
\label{probstate}

\subsection{Preliminaries}

In the basic form of OLDARP,  requests are issued dynamically to a server of unit capacity. Each request has source, destination, and release date. The server starts at an origin location and serves a request by picking up at the source and delivering at the destination.

We study competitive algorithms for variations of the OLDARP problem. We use standard terminology from competitive analysis.
In the context of OLDARP, an algorithm $\textsc{on}$ is considered \textit{online} if is learns about a request only at tis release time, whereas an algorithm is considered offline if it is aware of all requests at time 0 (i.e. the earliest time). We let \textsc{opt} denote the optimal offline algorithm. Given a sequence $\sigma = r_1, \ldots r_m$ of requests, we denote $\textsc{on}(\sigma)$ and $\textsc{opt}(\sigma)$ as the total revenue earned by $\textsc{on}$ and $\textsc{opt}$ respectively. \textsc{on} is \textit{c-competitive} if there exists $c > 0, b \ge 0$ such that \\

\centerline{$\textsc{on}(\sigma) \le c\cdot \textsc{opt}(\sigma) + b$}

\smallskip
	
We consider a modified version of the \textit{Online-Dial-A-Ride-Problem} on complete graphs. In this version, every request has a revenue and the goal is to serve requests such that the total revenue is maximized. The input is a complete graph where for every pair of nodes $u$ and $v$ there is a weight $w_{u, v} > 0$.  If for every edge $w_{u, v} = 1$ (i.e. the graph represents the unit metric space), we refer to the problem as ROLDARP.  If edge weights are varying, we refer to the problem as V-ROLDARP. One node in the graph, $o$, is designated as the origin, where a server is initially located. The input also includes a time limit $T > 2$ and a sequence of requests that is dynamically issued to the server. Each request is of the form $(s, d, t, r)$ where $s$ is the source node, $d$ is the destination, $t$ is the time the request is released, and $r$ is the revenue earned by the server for serving the request. We assume the earliest a request may be released is at time $t=0$. For each request, the server must decide whether to serve the request and if so, at what time. A request may not be served earlier than its release time and at most one request may be served at any given time. Once the server starts serving a request, it must serve the request until completion (i.e. preemption is not allowed).
The goal for the server is to maximize the total earned revenue of served requests. As a preprocessing step, we can remove any edge $(u, v)$ such that $w_{u, v} > T$, since no algorithm (either online or offline) can use this edge to serve a request.

We consider two variations of ROLDARP which we summarize below.

\begin{description}
\item[$\bullet$] original ROLDARP - Every edge in the graph has unit weight. Each request has a source, destination, release date, and revenue that is earned for serving the request. There is a global time limit before which requests must be served. The goal is to maximize the total revenue earned within the time limit.
\item[$\bullet$] V-ROLDARP - Edges in the graph have varying weights.
\end{description}

We first consider V-ROLDARP. The input to this problem is an undirected graph $G$ of $n \ge 2$ nodes where for every edge $(u, v)$ there is a weight $w_{u, v} >0 $; and there are at least two pairs of distinct edges $(u, v)$ and $(x, y)$ where $w_{u, v} \neq w_{x, y}$. Note that if $G$ contains exactly one edge, the weight of this edge must be greater than 1 (otherwise $G$ would be a unit distance graph).
We find that the offline version of V-ROLDARP is NP-Complete (see Section~\ref{offlinenpc}).
and that no online algorithm for V-ROLDARP can be competitive (Section~\ref{noncompete}). Note that for V-ROLDARP, any connected graph can be converted to a complete graph such that the pairwise distance between nodes of both graphs is equivalent. Therefore the proofs in Sections~\ref{offlinenpc} and~\ref{noncompete} also hold for non-complete graphs. Specifically, for two nodes $a$ and $b$ of a complete graph, where the weight of edge $(a, b)$ is $w$, we can create three nodes $a$, $b$, and $c$ and edges $(a, c)$ and $(c, b)$ for a non-complete graph where the weight of $(a, c) = k$ and the weight of $(c, b) = w-k$, for $k \ge 1$.   

Since no competitive algorithm for V-ROLDARP exists, we focus on ROLDARP and give a 2-competitive algorithm to solve this problem (see Section~\ref{compete}).

\section{Offline V-OLDARP is NP-Complete}
\label{offlinenpc}

We first show that the offline version of V-OLDARP, which we refer to as V-RDARP, is NP-Complete using
a reduction from the classical Traveling Salesperson Problem (TSP). We now formally define TSP and V-RDARP.

In TSP, the input is a value $k$ and a complete weighted graph $G$ of $n$ nodes where for each pair of nodes~$u$ and $v$ there is a weight $w_{u, v}$ of edge $(u, v)$. The problem asks: Is there a tour of cost at most $k$?

In V-RDARP, a server receives a sequence of requests and must decide for each request, whether to serve it or reject it. As in the online version of the problem, the input is a complete graph $G$ of $n$ nodes where for every pair of nodes $u$ and $v$ there is an edge with weight $w_{u,v}$; and there are at least two pairs of distinct edges $(u, v)$ and $(x, y)$ where $w_{u, v} \neq w_{x, y}$. One of the nodes is the origin $o$ where the server is initially located. The input also includes a time limit $T$, a goal revenue $R$, and a sequence of requests where each request is of the form $(s, d, t, r)$ where $s$ is the source, $d$ is the destination, $t$ is the release time, and $r$ is the revenue. A request may be satisfied at a time no sooner than its release time and at most one request may be satisfied during any duration. The server must start at the origin and choose a subset of requests to satisfy within time $T$ that earns total revenue at least $R$.

\subsection*{V-RDARP is NP-Complete}

\begin{description}

\item[$\bullet$] VDARP $\in$ NP.
A certificate is a sequence of requests $(s_1, d_1, t_1, r_1), (s_2, d_2, t_2, r_2), \ldots , (s_m, d_m, t_m, r_m)$ to satisfy and the start times $q_1, q_2, q_m$, respectively, for each request. To verify its correctness, we would ensure:

\begin{enumerate}
\item For every pair of adjacent requests $(s_i, d_i, t_i, r_i)$ and $(s_j, d_j, t_j, r_j), q_j \ge q_i + w_{s_i, d_i} + w_{d_i, s_j}$. In other words, a new request cannot be served before the server has completely served the previous request and has moved from the destination of the previous request to the start of the new request.
\item The finish time of the last request is at most $T$ (i.e. $q_m + w_{s_m, d_m} \le T$).
\item The sum of the revenues of each request is at least $R$ (i.e. $\Sigma_{i=1}^{m} r_i \ge R$).

\end{enumerate}

\


\item[$\bullet$] V-RDARP is NP-hard.

As previously mentioned, we reduce TSP to V-RDARP. Given an instance of TSP: a value $k$ and a complete weighted graph $G$ of $n$ nodes, we construct an instance of V-RDARP as described below. There is a TSP tour of cost $k$ in $G$ if and only if there is a set of requests that can be satisfied within time $T=k$ and earn total revenue $R = n$.

\smallskip

We construct an instance of V-RDARP as follows. Given the graph $G$ for TSP, we construct a graph $G^{\prime}$ for V-RDARP. $G^{\prime}$ is $G$ with an additional node $o$ (the origin) and edges from $o$ to every other node with weight 1 (i.e. for all nodes $u$, $w_{o, u} = 1$). For each edge $(u, v)$ in $G$, we have a request $(u, v, 1, 1 )$; we set the time limit $T = n$.

\smallskip

We claim that there is a TSP tour of cost $k$ in $G$ if and only if there is a set of requests with total revenue $n$ and service time at most $k+1$ in $G^{\prime}$:

\smallskip

First we show that if there is a set of requests for $G^{\prime}$ that earns revenue $R=n$ within time $k+1$, then  there is a TSP tour of cost $k$ in $G$. Suppose in $G^{\prime}$ there is a set of requests \

$(a, b, 1,1), (b, c, 1, 1),Ê(c, d, 1, 1) \ldots (v, a, 1, 1)$. Since the total revenue is $n$ and each request earns revenue 1, there must be $n$ requests and therefore each node must be the source and destination of exactly one request.  Satisfying this sequence of requests requires starting at $o$ and making a tour of the nodes of $G$. Since it takes 1 time unit to move from the origin to the first source (node $a$) and switching from one request to the next does not require any additional time, it must take $k$ time units to satisfy the set of requests.  Therefore the edges $(a, b), (b, c),Ê(c, d) \ldots (v, a)$ must make a tour of cost at most $k$ in $G$.

\smallskip

Now we show that if there is a TSP tour of cost $k$ in $G$ then there is a sequence of requests with total revenue $n$ that can be fulfilled within time $k+1$ in $G^{\prime}$.Ê If there are $n$ nodes in $G$, then a tour of $G$ will consist of $n$ edges. Suppose a set of $n$ edges $(a, b), (b, c), (c, d) \ldots (v, a)$ makes up the TSP tour in $G$, then we can fulfill the requests $(a, b, 1, 1), (b, c, 1, 1), (c, d, 1, 1) ... (v, a, 1, 1$) earning total revenue $R = n$. Since the source of each request is the destination of the previous request, switching from one request to the next does not require any time. Since $G$ contains a tour of cost $k$ and moving from the origin to the source takes time 1,  the total time to fulfill the requests is $T=k+1$.

\end{description}

\section{Non-Competitiveness of V-ROLDARP}
\label{noncompete}

In this section, we prove a negative result regarding the non-competitiveness of the V-ROLDARP problem. Whereas many non-competitive results assert that no deterministic online algorithm can be competitive for \textit{all} inputs to a problem, we make a stronger assertion: no deterministic online algorithm can be competitive for \textit{any} input to the V-ROLDARP problem.  We prove this by way of contradiction.  Let $\sigma$ denote a sequence of requests, \textsc{opt} denote an optimal offline algorithm, \textsc{on} denote a \textit{c-competitive} online algorithm (for $c \ge 1$), $\textsc{opt}(\sigma)$ denote the revenue earned \textsc{opt}, and $\textsc{on}(\sigma)$ denote the revenue earned by \textsc{on}.  We assume (on the contrary) that $\textsc{opt}(\sigma) \le c \cdot \textsc{on}(\sigma) + b$ for some $b > 0$.

We will show that for any graph $G$ of $n \ge 2 $ nodes and any time limt $T > 2$, an adversary can construct $\sigma$ such that \textsc{opt} earns $b$ more revenue than \textsc{on}, i.e. $\textsc{opt}(\sigma) - \textsc{on}(\sigma) = b+ \epsilon$ for $\epsilon > 0$.

The main idea is as follows: for \textsc{on} to be competitive it must accept some request. As soon as \textsc{on} accepts a request, the adversary will generate a new request with a arbitrarily high revenue that can be accepted by \textsc{opt} but (due to the time limit) not by \textsc{on}.

Let $(u, v)$ be an edge in $G$ with weight $1 < w_{u, v} < T$.  As stated in the problem statement, we can assume w.l.o.g. that $w_{u, v} < T$; also since edge weights are varying, if $G$ contains at least 2 edges, then there must be some edge such that $1 < w_{u, v}$.  Therefore such an edge must exist. The adversary will generate a request $r_1 = (u, v, T-w_{u, v}-1, b+\epsilon)$. If \textsc{on} rejects, then the adversary will generate no more requests. \textsc{on} will earn revenue 0 while \textsc{opt} will earn revenue $b+\epsilon$. If \textsc{on} accepts the request, then at the next time unit the adversary will generate a new request $r_2 = (u, v, T-w_{u, v}, 2(b + \epsilon))$. In the best case for \textsc{on}, the origin will be node $u$, so \textsc{on} will finish serving $r_1$ at time $T-1$. \textsc{on} will not arrive at $u$ to serve $r_2$ before time $T$ and therefore not be able to serve $r_2$. \textsc{opt} will reject $r_1$ and serve $r_2$ starting at time $T-w_{u, v}$ (the release time) and finishing at time $T$.

\subsection{Non-Competitiveness of V-ROLDARP with Preemeption}

We now consider a variation of V-ROLDARP where the server has the option preempt a request that it is currently serving; in other words, the server may change its mind about serving a request.  In particular, while serving a request, the server may decide to quit the current request and switch to a newly released request. In this setting the server may or may not be required to spend time to move to either the source or destination of the current request. We show that even in the case where the server is not required to spend this time, no online algorithm can be competitive. Furthermore, the non-competitiveness holds for any graph that contains two nodes $v$ and $y$ such that the hop distance between $v$ and $y$ is at least 3.

As before let $(u, v)$ be an edge in $G$ with weight $1 < w_{u, v} < T$, which we know must exist. Let $(x, y)$ be an edge such that the hop distance between $v$ and $y$ is at least 3. The adversary will generate a request $r_1 = (u, v, T-w_{x, y}-1, b+\epsilon)$. If \textsc{on} rejects, then the adversary will generate no more requests. \textsc{on} will earn revenue 0 while \textsc{opt} will earn revenue $b+\epsilon$. If \textsc{on} accepts the request, then at the next time unit the adversary will generate a new request $r_2 = (x, y, T-w_{x, y}, 2(b + \epsilon))$. If \textsc{on}'s server quits serving $r_1$ when $r_2$ is released, it must move to $x$ from either $u$ or $v$.  In the best case for \textsc{on}, $w_{u, x} =1 $ (or $w_{v, x} = 1$) so the earliest the server can arrive at $x$ is at time $T-w_{x, y}+1$, which is too late to serve $r_2$. On the other hand, \textsc{opt} will reject $r_1$ and serve $r_2$ starting at time $T-w_{x, y}$ (the release time) and finishing at time $T$. \textsc{on} will earn revenue $0$ while \textsc{opt} will earn revenue $b + \epsilon$. If \textsc{on}'s server does not quit serving $r_1$, \textsc{on} will earn revenue $b$ while \textsc{opt} will earn revenue $v + \epsilon$.

 \section{ROLDARP on a Complete Graph with Unit Edges}

\begin{algorithm}\caption{Given graph $G$ and time limit $T$}
\begin{algorithmic}[1]

\label{algo}

\IF {$T$ is even}

 \STATE At every even time, determine which released request earns the greatest revenue and move to the source location of this request.  Denote this request as $r$. If no unserved requests exist, do nothing until the next odd time.

 \STATE At every odd time, complete request $r$ from the previous step.

 \ENDIF

 \IF {$T$ is odd}

\STATE At time 0, do nothing.

 \STATE At every odd time, determine which released request earns the greatest revenue and move to the source location of this request.  Denote this request as $r$. If no unserved requests exist, do nothing until the next even time.

 \STATE At every even time, complete request $r$ from the previous step.
 \ENDIF

\end{algorithmic}
\end{algorithm}

 \label{compete}

In this section, we provide an online algorithm, Greatest Revenue First (\textsc{grf}) that is 2-competitive for the ROLDARP. Specifically, given request sequence $\sigma$, if $\textsc{opt}(\sigma)$ denotes the optimal revenue earned from $\sigma$ and $\textsc{grf}(\sigma)$ denotes the amount of revenue earned by $\textsc{grf}$ from $\sigma$, we show:

\begin{eqnarray}
\label{compeqn}
\textsc{opt}(\sigma) \le 2 \cdot \textsc{grf}(\sigma) + b
\end{eqnarray}

\noindent
where $b$ is the revenue of the last request. We first show that for any graph $G$ with $n \ge 2$ nodes and a time limit $T > 2$ no online algorithm can avoid the $b$ additive factor of the equation in~\ref{compeqn}.  In particular an adversary can generate a request sequence such that no online algorithm can serve the last request of the sequence.

To do this, the adversary will first generate a request $r_1 = (u, v, T-w_{u, v}-1, b_1)$. If the online algorithm rejects the request, the adversary will generate no more requests. An optimal offline algorithm will accept $r_1$ and earn revenue $b_1$. If the online algorithm accepts the request, then at the next time unit the adversary will generate a new request $r_2 = (u, v, T-w_{u, v}, b_2)$. In the best case for the online algorithm, the origin will be node $u$, so the server will finish serving $r_1$ at time $T-1$. The server will not arrive at $u$ to serve $r_2$ before time $T$ and therefore not be able to serve $r_2$. The server for an optimal offline algorithm will reject $r_1$ and serve $r_2$ starting at time $T-w_{u, v}$ (the release time), finishing at time $T$, and earning $b_2$.

 Algorithm~\ref{algo} describes the \textsc{grf} algorithm. The main idea is that for every time unit for which there is some unserved request, \textsc{grf} either moves to the source location of the request with the highest revenue or completes a previous request.

\subsection{Greatest Revenue First is 2-competitive}

We now prove that \textsc{grf} is 2-competitive. Let $\textsc{alg}$ denote an algorithm that serves requests, then $VAL(\textsc{alg})$ denotes the sum of all revenues of the requests served by $\textsc{alg}$. Let $v_{last}$ refer to the revenue of the last request served by \textsc{opt}.  To prove that \textsc{grf} is 2-competitive, we must prove the following equation:

\begin{equation}
2 \cdot VAL(\textsc{grf}) + v_{last} \ge VAL(\textsc{opt})
\label{rvsopt}
\end{equation}

To prove this equation, we consider another algorithm \textsc{max}. We define \textsc{max} such that at every time unit except $T-1$, \textsc{max} serves the request with the greatest revenue regardless of the source node of the request. At time $T-1$, \textsc{max} does nothing. Note that \textsc{max} may not coincide with the request set of a feasible algorithm. In other words, given the input graph, request sequence, and time limit, \textsc{max} may fulfill a set of requests that \textit{no} algorithm can fulfill. For example, suppose the origin is some node $o$. Suppose at time 0, a request with maximal revenue is released with source $s_0 \neq o$ and destination $d_0$. No algorithm can complete this request in the time slot from 0 to 1, but we assume that \textsc{max} does. Thus, by the construction of \textsc{max}, the following equation holds:

\begin{equation}
VAL(\textsc{max}) \ge VAL(\textsc{OPT}) - v_{last}
\label{rvopt}
\end{equation}

We will show that:
\begin{equation}
2 \cdot VAL(\textsc{grf})\ge VAL(\textsc{max})
\label{rvsmax}
\end{equation}

A proof of~(\ref{rvsmax}) will immediately prove~(\ref{rvsopt}).

\begin{table}
\caption{$T=6$, \textsc{grf} values given $WLOG$}
\centering
\begin{tabular}{c c c}
\hline
\hline
$\tau$ & \textsc{grf} & \textsc{max} \\[0.5ex]
\hline
0 & 0 & $v_0$\\
1 & $v_0$ & $v_1$\\
2 & 0 & $v_2$\\
3 & $v_1 = max(v_1,v_2)$ & $v_3$\\
4 & 0 & $v_4$\\
5 & $v_3 = max(v_2,v_3,v_4)$ & 0\\
[1ex]
\hline
\end{tabular}
\label{table:nonlin}
\end{table}

We now prove equation~(\ref{rvsmax}). For this proof we use the terminology ``algorithm $A$ \textit{serves} (or \textit{has served}) request $r$ at time $t$" to indicate that $A$ \textit{begins} serving $r$ at time $t$ and \textit{completes} serving $r$ at time $t+1$.

We assume without loss of generality, that there exists enough requests such that at every time unit \textsc{grf} and \textsc{max} have a request to serve.

Let $r_0, r_1, r_2, \ldots r_{T-2}$ denote the requests served by \textsc{max} at times $0, 1, 2, \ldots T-2$ earning revenues $v_0, v_1, v_2, \ldots v_{T-2}$. We consider two cases based on the parity of $T$.

Case 1: $T$ is even.

At $t=0$, \textsc{grf} and \textsc{max} determine that $v_0$ is the greatest revenue. At $t=0$ \textsc{max} fulfills $v_0$ and at $t=1$ \textsc{grf} fulfills $v_0$.

We now show that for every odd time $t \neq 1$, if \textsc{max} serves $r_{t-1}$ and $r_{t-2}$ at times $t-1$ and $t-2$, earning total revenue $v_{t-1} + v_{t-2}$, then at time $t$ \textsc{grf} earns revenue at least $\max\{v_{t-1}, v_{t-2}\}$. Note that when $T$ is even \textsc{grf} serves requests only at odd times. We show that at time $t-1$, when \textsc{grf} decides which request to complete at time $t$, both $r_{t-1}, r_{t-2}$ are available requests.

Consider $r_{t-1}$ and $r_{t-2}$. Since \textsc{max} has served them at $t-1$ and $t-2$, they must have been released by times $t-1$ and $t-2$ respectively. The only way that they would not be available requests for \textsc{grf} at time $t-1$ is if \textsc{grf} has already completed them. However, this is not possible because if \textsc{grf} completes a request at some time $\tau$, \textsc{max} must complete that request by time $\tau -1$ at the latest.

Case 2: $T$ is odd.

We now show that for every even time $t \neq 0$, if \textsc{max} serves $r_{t-1}$ and $r_{t-2}$ at times $t-1$ and $t-2$, earning total revenue $v_{t-1} + v_{t-2}$, then at time $t$ \textsc{grf} earns revenue at least $\max\{v_{t-1}, v_{t-2}\}$. Note that when $T$ is odd \textsc{grf} serves requests only at even times. We show that at time $t-1$, when \textsc{grf} decides which request to complete at time $t$, both $r_{t-1}, r_{t-2}$ are available requests.

Consider $r_{t-1}$ and $r_{t-2}$. Since \textsc{max} has served them at $t-1$ and $t-2$, they must have been released by times $t-1$ and $t-2$ respectively. The only way that they are not available requests for \textsc{grf} at time $t-1$ is if \textsc{grf} has already completed them. However, this is not possible because if \textsc{grf} completes a request at some time $\tau$, \textsc{max} must complete that request by time $\tau -1$ at the latest.


\medskip

Now, let $v^{\prime}_t$ denote the revenue earned by \textsc{grf} at time $t$. Then, for all times $t$ in which \textsc{grf} earns revenue:

\begin{eqnarray}
v_t^{\prime} \ge& \max\{v_{t-1},v_{t-2}\} \\
2 \cdot v_t^{\prime} \ge& v_{t-1}+v_{t-2}
\label{allt}
\end{eqnarray}

Let $r^{\prime}_t$ denote the request served by \textsc{grf} at time $t$. Since~(\ref{allt}) holds for all $r^{\prime}_t$ served by \textsc{grf}, we have:

\begin{equation}
2 \cdot VAL(\textsc{grf}) \ge VAL(\textsc{max})
\end{equation}

Then, from~(\ref{rvopt}), we have

\begin{eqnarray}
2 \cdot VAL(\textsc{grf}) \ge& VAL(\textsc{OPT}) - v_{last} \\
2 \cdot VAL(\textsc{grf}) + v_{last} \ge& VAL(\textsc{OPT})
\end{eqnarray}


\begin{thebibliography}{4}

\footnotesize

\bibitem{ausiello} G. Ausiello, E. Feuerstein, S. Leonardi, L. Stougie, and M. Talamo. \em{Algorithms for the On-Line Travelling Salesman}.

\bibitem{ausiello2}G. Ausiello, V. Bonifaci, and L. Laura. \em{The On-Line Asymmetric Traveling Salesman Problem}.

\bibitem{jaillet2} P. Jaillet and X. Lu. \em{Online Traveling Salesman Problems with Flexibility.}


\bibitem{krumke} On Minimizing the Maximum Flow Time in the Online Dial-a-Ride Problem. \em{ Networks}, vol. 44, pp. 41-46, 2004.

\bibitem{lipmann} M. Lipmann, X. Lu, W. E. de Paepe, R. A. Sitters, and L. Stougie. \em{ On-line Dial-a-Ride Problems Under Restricted Information Model}. Algorithmica, 40:319 329, 2004.

\bibitem{stougie} On-Line Single-Server Dial-a-Ride Problems.


\end{thebibliography}
\end{document}